\documentclass[aip,pop,reprint]{revtex4-1}

\usepackage{graphicx}
\usepackage{dcolumn}
\usepackage{bm}
\usepackage{float}
\usepackage{subfigure}
\usepackage{epstopdf}

\begin{document}
\title{Extreme Nonlinear Dynamics in Vacuum Laser Acceleration with a Crossed Beam Configuration}
\author{A.P.L.Robinson}
\affiliation{Central Laser Facility, STFC Rutherford-Appleton Laboratory, Didcot, OX11 0QX, UK}
\author{K.Tangtartharakul}
\affiliation{Department of Mechanical and Aerospace Engineering, University of California at San Diego, La Jolla, CA 92093, USA}
\author{K.Weichman}
\affiliation{Department of Mechanical and Aerospace Engineering, University of California at San Diego, La Jolla, CA 92093, USA}
\author{A.V.Arefiev}
\affiliation{Department of Mechanical and Aerospace Engineering, University of California at San Diego, La Jolla, CA 92093, USA}
\date{\today}
\begin{abstract}
A relatively simple model problem where a single electron moves in two relativistically-strong obliquely intersecting plane wave-packets is studied using a number of different numerical solvers.  It is shown that, in general, even the most advanced solvers are unable to obtain converged solutions for more than about 100~fs in contrast to the single plane-wave problem, and that some basic metrics of the orbit show enormous sensitivity to the initial conditions.   At a bare minimum this indicates an unusual degree of non-linearity, and may well indicate that the dynamics of this system are chaotic.
\end{abstract}

\maketitle

\section{Introduction}
Since the development of Chirped Pulse Amplification lasers \cite{danson_vulcan_2004,hernandez-gomez_vulcan_2010,tang_optical_2008,hooker_improving_2011}, the field of ultra-intense laser-matter interactions has grown considerably.  Initially this technology allowed the 
development of TW-scale lasers that breached the 10$^{18}$Wcm$^{-2}\mu$m$^2$, but subsequent progress has lead to the construction of 10PW scale systems \cite{hernandez-gomez_vulcan_2010}, with
100PW systems under development.  The field now spans a large number of sub-topics including laser wakefield acceleration of electrons \cite{mangles_monoenergetic_2004}, laser-driven ion acceleration \cite{gaillard_increased_2011}, laser-
driven x-ray \cite{kneip_observation_2008} and neutron sources, advanced inertial fusion concepts such as Fast Ignition \cite{m.tabak_review_2005}, studies of both Warm Dense Matter and Hot Dense Matter \cite{d.j.hoarty_observations_2013}, radiation reaction studies, and even probing QED physics \cite{heinzl_observation_2006,heinzl_exploring_2009}.  It is likely that the latter topics in that list will become more dominant as multi-PW facilities become fully operational in the following years.  Numerical simulation codes, particularly Particle-in-Cell (PIC) codes \cite{birdsall_plasma_2018,pukhov_strong_2003}, have been instrumental in driving the field forward, both in terms of interpreting experiments and in making predictions that have motivated crucial experimental work.  Perhaps the best known example of PIC's predictive capabilities is that of Pukhov and Meyer-ter-Vehn's prediction \cite{pukhov_laser_2002} of the `bubble regime' of laser wakefield acceleration, which was later validated by three different research groups \cite{mangles_monoenergetic_2004}. 

The PIC algorithm is itself dependent on a number of algorithms, some of which were developed separately, such as the Yee FDTD method \cite{kane_yee_numerical_1966} for numerical electromagnetics.  
Importantly this includes a  `particle-pusher' algorithm which advances the macroparticles position and momentum.  The quality and capability of any individual PIC code will depend on the set of algorithms chosen for these different components.  A common choice for the particle-pusher is the Boris method \cite{birdsall_plasma_2018}.  The Boris method is a second order accurate leapfrog-type method that is centred in time.  It is a method that has enjoyed considerable success, and which has been employed in a number of different PIC codes including EPOCH \cite{arber_contemporary_2015}.  Developing higher order versions of the Boris method is a non-trivial proposition, and it is has been questioned whether or this endeavour would actually yield any serious benefits to laser-plasma or accelerator science \cite{higuera_structure-preserving_2017,londrillo_2010}.

In the past few years however it has been recognized that the Boris method has at least one serious defect, namely that constant motion is not maintained in the case of uniform crossed ${\bf E} \ne 0$ and ${\bf B} \ne 0$ fields (for the choice of particle velocity for which a force-free scenario is obtained).  This was first identified by Vay \cite{vay_simulation_2008}, who proposed a variation on the Boris pusher that resolved this issue.  Later Higuera and Cary \cite{higuera_structure-preserving_2017} proposed an algorithm that both solved the issue of the ${\bf E} \times {\bf B}$ velocity and which also preserved phase-space volume (unlike Vay's method).  Alongside these developments, Arefiev also showed that considerable care needs to be taken in setting the time step when integrating the orbits of an electron in a relativistic laser pulse.  Altogether these developments underline how the particle-pusher problem needs careful study to ensure that particle-pusher algorithms can be trusted when employed to study the strongly relativistic and highly complex configurations encountered in ultra-intense laser-matter problems.  

Despite these developments the methods of Vay and Higuerra-Cary are still only second order accurate methods.  For problems where the overall behaviour of the system is quite 'regular' this means that they will be quite adequate in the majority of cases.  What has not been given so much consideration is whether the dynamics can always be assumed to be sufficiently `regular'.  Some researchers have pointed out that some laser interaction problems will have a `stochastic' nature \cite{z.-m.sheng_stochastic_2002,sheng_efficient_2004,meyer-ter-vehn_electron_1999}, this terminology appears to actually mean that the dynamics are {\it chaotic}\cite{strogatz_nonlinear_2015,ott_coping_1994,handfinch}.  If the Lyapunov time is larger than the time-scale of interest then this is not a problem for numerical simulation.  However if the Lyapunov time becomes much shorter than the time-scale of interest then the  ability to predict future states of the system will be highly limited even with very sophisticated numerical solvers.

In this paper we present a relatively simple test problem for a single electron :  two plane EM Gaussian wave-packets that cross at an oblique angle and which are $\pi$ out of phase.  The electron is initially at rest and which sits `off-axis' by a fraction of the vacuum wavelength.  To the best of the authors' knowledge this problem does not have an analytic solution.  We have studied the ability of a number of leapfrog pushers, RK4 method, and more sophisticated adaptive algorithms to solve the electron orbits in this problem.  We have found that, in general, all of these solvers are only able to obtain converged orbits for a fraction ($<$20\%) of the total problem duration (100-200~fs out of 1~ps).  Complete converged orbits are only obtained in a few cases, and usually only the RK4 method (or better) is able to do this.  A survey of the sensitivity to initial conditions was carried out, and it was found that there are regions of parameter space which exhibit extreme sensitivity to initial conditions.  This indicates that this problem, however simple it may seem, in fact is {\it chaotic} in nature, as expected given earlier studies \cite{z.-m.sheng_stochastic_2002,sheng_efficient_2004,meyer-ter-vehn_electron_1999}, however in this case it would appear that the chaotic dynamics is severly problematic for numerical integration.  We suggest that this may have important ramifications for both Vacuum Laser Acceleration (VLA) \cite{hartemann_chirped-pulse_1999,thevenet_vacuum_2016,plettner_proof--principle_2005,troha_vacuum_1999,robinson_interaction_2018} and Direct Laser Acceleration (DLA) \cite{pukhov_particle_1999,naseri_channeling_2012,arefiev_beyond_2016,robinson_breaking_2017,arefiev_spontaneous_2016,huang_characteristics_2016,zhang_synergistic_2015,willingale_surface_2013,robinson_generating_2013}.

%###################################################
%##### Description of the Problem.
%###################################################
\section{Description of Model Problem}
\label{model}
We consider a problem where two relativisitically-strong plane EM wave-packets intersect obliquely.  We want to study the relativistic motion of an electron that is initially at rest.
This can be described by the following formulae for the electric fields of the incident waves:

\begin{eqnarray}
&&{\bf E} = {\bf E}_1 + {\bf E}_2, \\
&&{\bf E}_1 = E\cos\psi_1f_{env,1}\left[ -\sin(\theta_{cb}/2),\cos(\theta_{cb}/2),0\right], \\
&&{\bf E}_2 = E\cos\psi_2f_{env,2}\left[ \sin(\theta_{cb}/2),\cos(\theta_{cb}/2),0\right],
\end{eqnarray}

\begin{figure}
    \centering
    \includegraphics[width=\columnwidth]{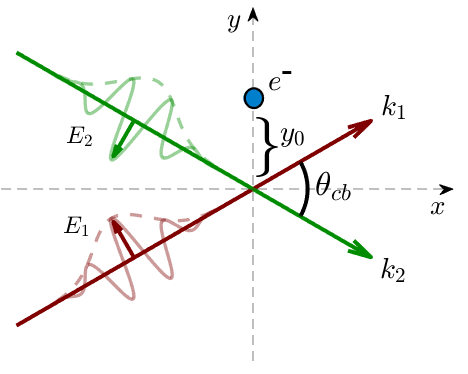}
    \caption{Schematic of the simulation set-up showing key parameters.}
    \label{sketch}
\end{figure}

\noindent where $\psi_1 = {\bf k_1.r} - \omega_L{t} + \phi_1$, $\psi_2 = {\bf k_2.r} - \omega_L{t}+\phi_2$, ${\bf k}_1 = [\cos(\theta_{cb}/2),\sin(\theta_{cb}/2),0]$, ${\bf k}_2 = [\cos(\theta_{cb}/2),-\sin(\theta_{cb}/2),0]$.  For the envelope functions, we use $f_{env} = \exp(-(\psi/k_L + 5c\tau_L)^2/(2c\tau_L))$.  There are corresponding magnetic fields in the z-direction.  This corresponds to two intersecting plane wave-packets that are aligned obliquely to the $x$-axis with the E-field polarized in the $xy$-plane in each case.  The angle between the wavevectors of the two wave-packets is $\theta_{cb}$.  For our baseline problem we consider the case where $E = 5\omega_Lm_ec/e$ (i.e. each plane wave-packet has $a_0 =$ 5), $\theta_{cb}= $40$^\circ$, $\lambda_L =$1 $\mu$m, and $\tau_L = $20~fs.  The two wavepackets are $\pi$ out of phase, i.e. $\phi_1 =$ 0, $\phi_2 = \pi$. The electron is initially at rest at $x =$0,$z = $0, and $y = y_0$.  A schematic of the problem is shown in fig.\ref{sketch}.

Since this problem is quite close to that considered previously \cite{z.-m.sheng_stochastic_2002,sheng_efficient_2004,meyer-ter-vehn_electron_1999}, we should expect that chaotic dynamics are likely to be encountered.  A very significant difference with earlier studies is that the value of the normalized vector potential in this case is significantly larger ($a >$ 5 here).  However since Mendonca's \cite{mendonca_1983} criterion is $a_1a_2 >$1/16 we expect that chaotic dynamics will only be more prevalent in this problem.

%###################################################
%##### Analysis Phase One
%###################################################
\section{Analysis with Standard Algorithms}
\label{standard}
%Solvers
In the first part of our study we have used the following solvers : (i) the Boris pusher \cite{birdsall_plasma_2018}, (ii) the Vay pusher\cite{vay_simulation_2008}, (iii) the Higuera-Cary pusher\cite{higuera_structure-preserving_2017}, and (iv) the 4th order Runge-Kutta (RK4) algorithm \cite{numrecipes}, to study this problem.  Note that (i)--(iii) are formally 2nd order algorithms (although they differ in their treatment of the ${\bf E} \times {\bf B}$ velocity) , and only (iv) is formally 4th order.  These were applied to study the baseline case (case 1).  We shall not re-state the details of these here, and we refer the reader to the given references for further details.  We have tested and checked our implementations, in particular by testing that they reproduce the motion in a single plane wave-packet.  The baseline numerical integration is carried out over 18000 time steps with $\Delta{t} = $0.05~fs.  To examine convergence, the time step is multiplied by a factor $1/M$, and the total number of time steps by $M$ in order to keep the total duration of the integration constant.  In general, we regard two trajectories as being converged if the variables in question are within 5\% of one another.  All of these solvers reproduce the analytic prediction for the single plane-wave problem with $M$ = 1 and the solutions of each solver are practically identical.  

\begin{figure}
\includegraphics[width=\columnwidth]{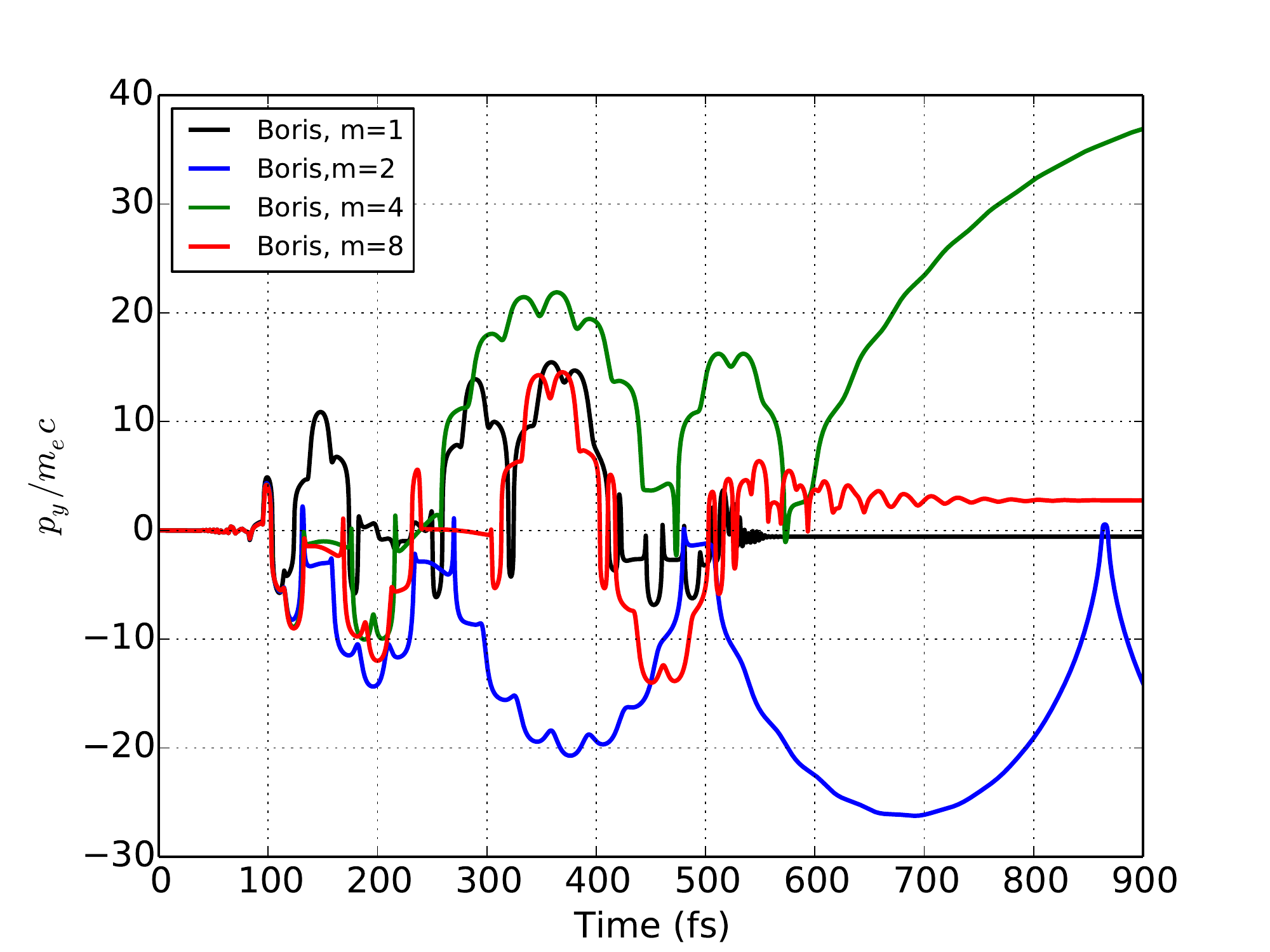}
\caption{\label{fig:figure1}The results from the Boris Pusher for the baseline case.  Value of $M$ for each line is shown in the legend.  Solution shows no sign of convergence with increasing $M$.}
\end{figure}

%Results
For each solver we obtained solutions of $M =$1,2,4, and 8.  The results for the Boris pusher, in terms of $p_y$ are shown in fig.\ref{fig:figure1}.  By following sequence of cases, we can see that the solution is not converging.

\begin{figure}
\includegraphics[width=\columnwidth]{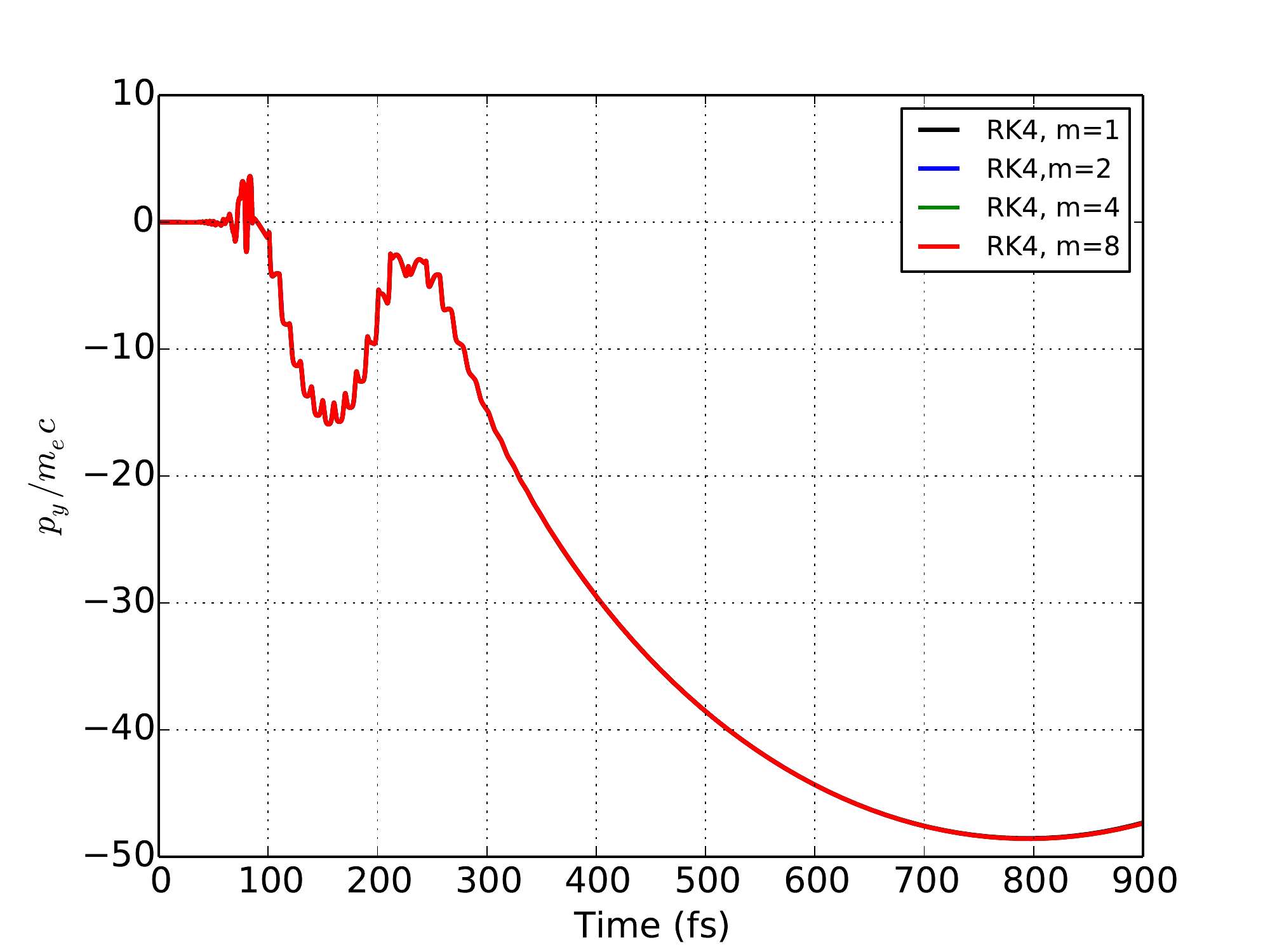}
\caption{\label{fig:figure2}The results from the RK4 algorithm for the baseline case.  Value of $M$ for each line is shown in the legend. All four curves overlap almost perfectly, indicating extremely rapid convergence.}
\end{figure}

The behaviour of the Boris pusher is in sharp contrast with the RK4 algorithm.  The results of the RK4 algorithm, also in terms of $p_y$, are shown in fig.\ref{fig:figure2}.  Here the four curves almost perfectly overlap, showing clearly that there has been very good convergence, and that it has happened very rapidly.

\begin{figure}
\includegraphics[width=\columnwidth]{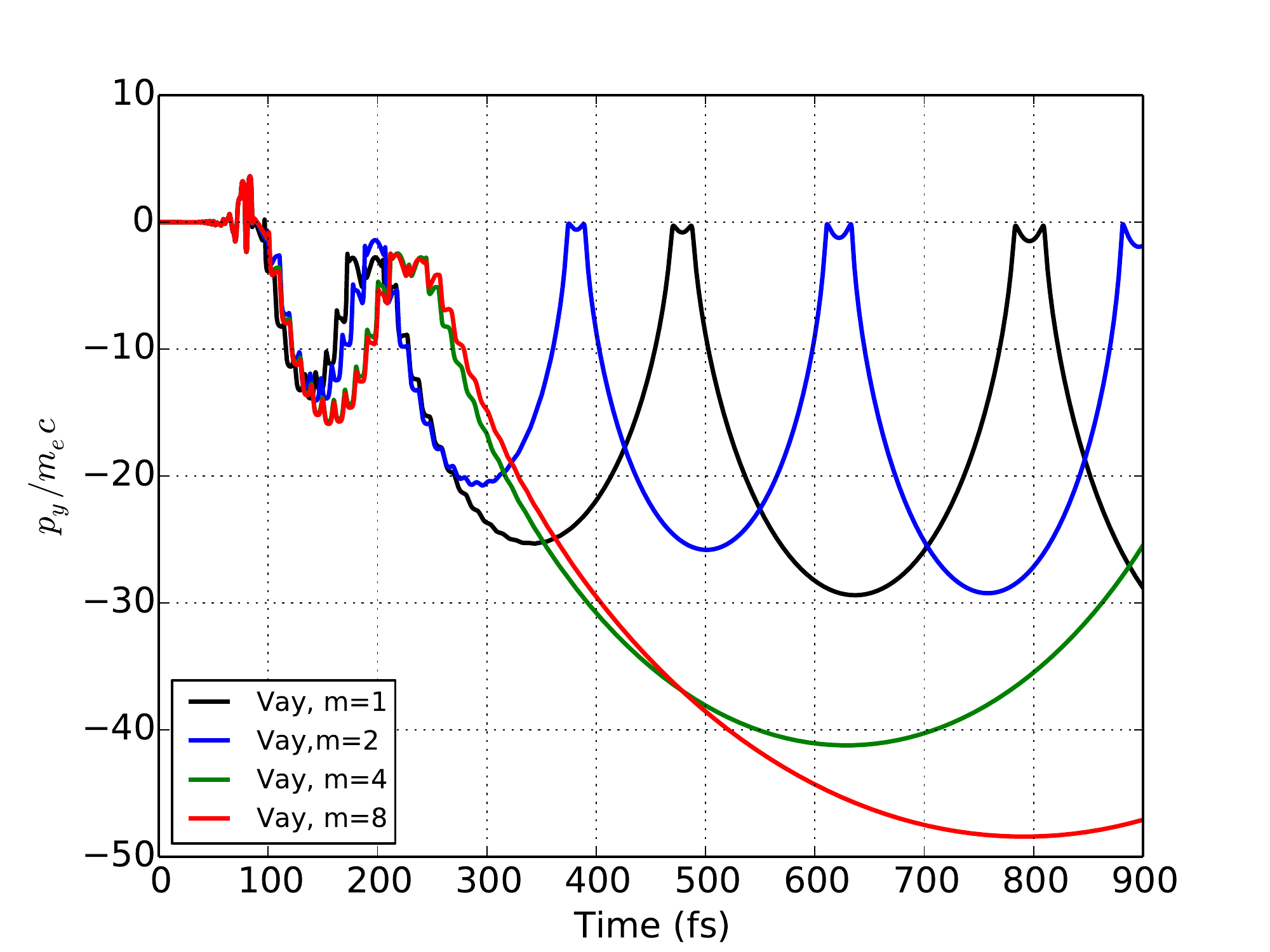}
\caption{\label{fig:figure3}The results from the  Vay Pusher for the baseline case.  Value of $M$ for each line is shown in the legend. }
\end{figure}

\begin{figure}
\includegraphics[width=\columnwidth]{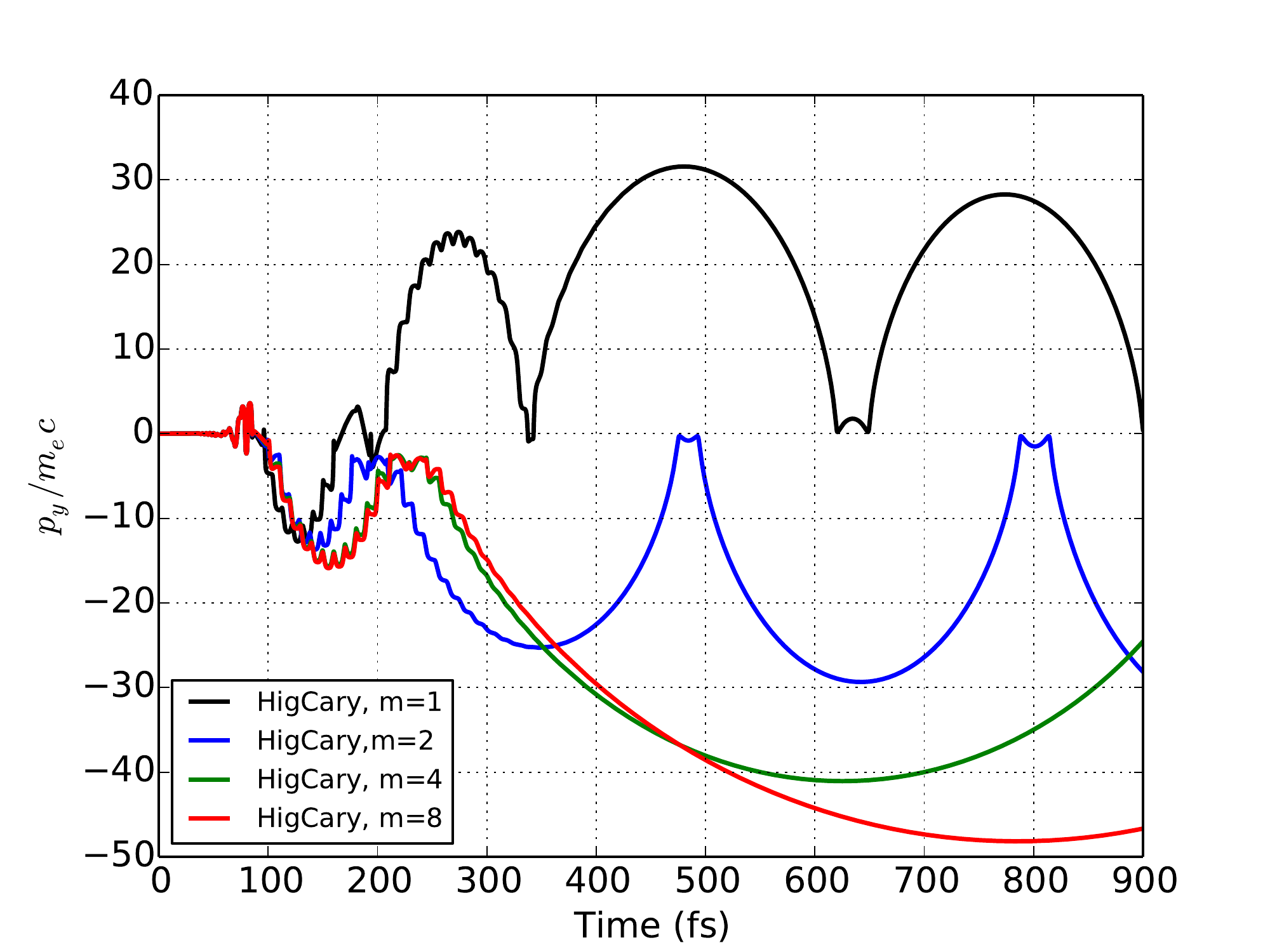}
\caption{\label{fig:figure4}The results from the Higuera-Cary pusher for the baseline case.  Value of $M$ for each line is shown in the legend. }
\end{figure}

\begin{figure}
\includegraphics[width=\columnwidth]{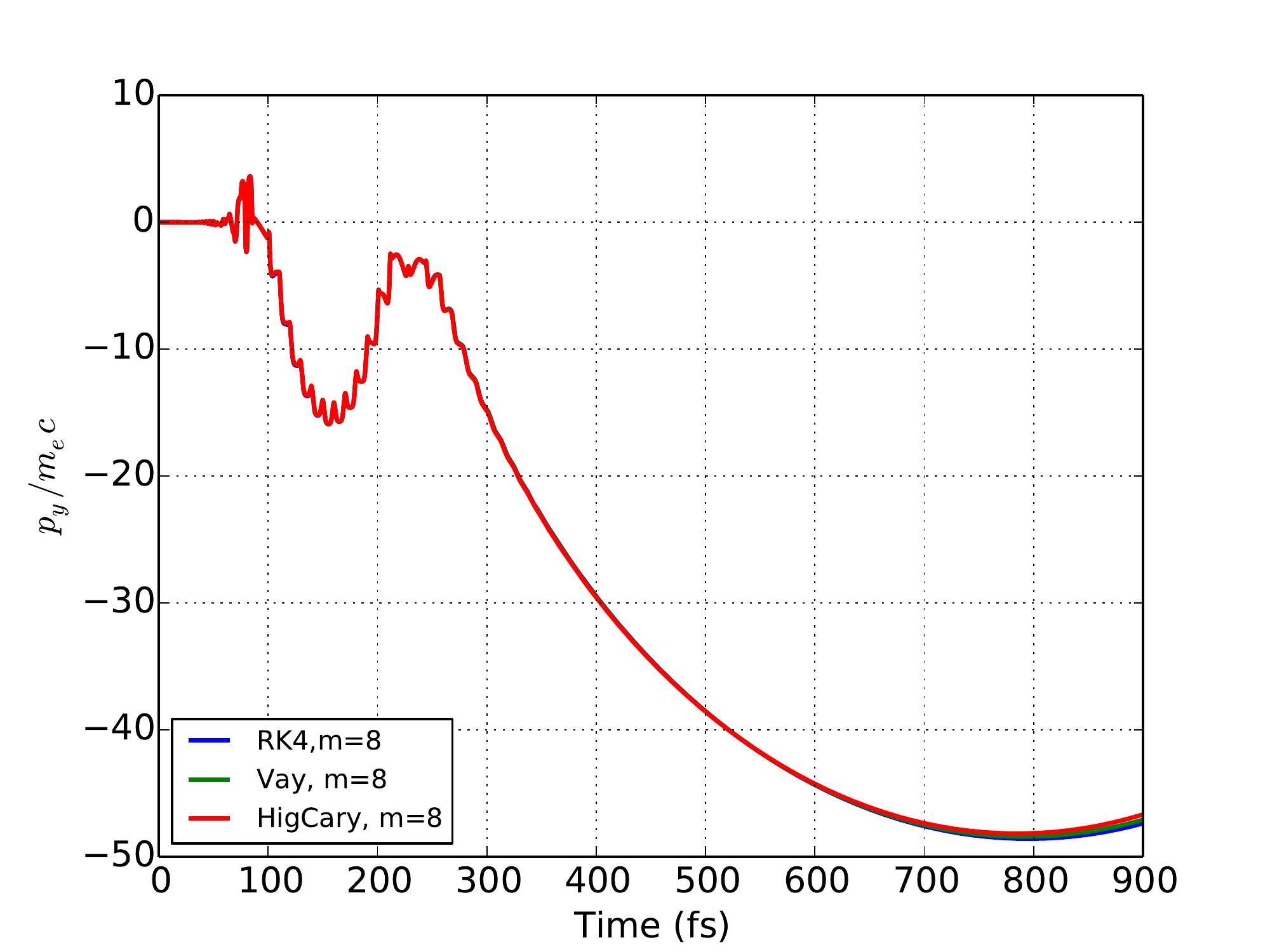}
\caption{\label{fig:figure5}Comparison of the solutions from the Vay, Higuera-Cary, and RK4 pushers for $M =$8, showing that, in the $M =$8 case, convergence is obtained. }
\end{figure}

The behaviour of both the Vay and Higuera-Cary pushers are shown in fig.s~\ref{fig:figure3} and \ref{fig:figure4}.  By comparing fig.s~\ref{fig:figure3} and \ref{fig:figure4} to fig.~\ref{fig:figure2} we can see that, when $M =$8, both the Vay and the Higuera-Cary pushers come very close to the solution obtained by the RK4 algorithm.  This should lead to confidence in the solution obtained by the RK4 solver.  It is clearly good that both the Vay and Higuera-Cary pushers are able to eventually reach this solution, however the rate of convergence is rather slow, and it requires that one adopts a very small time ($M =$8) time step.   In figure \ref{fig:figure5} we directly compare the Vay, Higuera-Cary, and RK4 solutions for $M =$8.  As can be seen they all lie extremely close to one another, showing that the Vay and Higuera-Cary solvers are able to approach the RK4 solution, whereas the Boris solver cannot for $M \le 8$. 

In the second part of our study we extended this to multiple cases to see if these findings reflected a general trend.  As is evident from fig.s \ref{fig:figure1}--\ref{fig:figure4}, even when convergence is not obtained over the entire 900~fs, convergence in fact can occur over a time period that is a fraction of the total duration of the problem.  When extending the study we instead looked at the fraction of the problem duration over which convergence was obtained (instead of whether or not {\it total} convergence was obtained).  The results are summarized in table \ref{table1}, which shows the convergence obtained for each case as a percentage of the total problem duration (900~fs), and for each solver tried.  The special cases of the convergence obtained by the Vay and Higuera-Cary pushers in the baseline case are noted by an asterisk.    

\begin{table}
\begin{center}
\begin{tabular}{|c|c|c|c|c|}
\hline
Case & Boris & Vay & Hig.-Cary & RK4 \\
\hline
1.$a_0 =$5, $y_0 = \lambda/4$, $\theta_{cb} = $ 40$^\circ$ & 14.7 & 100* & 100* & 100 \\
\hline
2.$a_0 =$5, $y_0 = \lambda/2$, $\theta_{cb} = $ 40$^\circ$ & 11.1 & 11.0 & 9.1 & 12.1 \\
\hline
3.$a_0 =$10, $y_0 = \lambda/2$, $\theta_{cb} = $ 40$^\circ$ & 12.8 & 13.3 & 12.9 & 100.0 \\
\hline
4.$a_0 =$10, $y_0 = \lambda/4$, $\theta_{cb} = $ 40$^\circ$ & 10.2 & 10.2 & 11.3 & 17.8 \\
\hline
5.$a_0 =$5, $y_0 = \lambda/8$, $\theta_{cb} = $ 40$^\circ$ & 10.9 & 11.6 & 11.6 & 17.6\\
\hline
6.$a_0 =$10, $y_0 = \lambda/8$, $\theta_{cb}= $ 40$^\circ$ & 10.8 & 11.9 & 11.9 & 17.1 \\
\hline
7.$a_0 =$5, $y_0 = \lambda/4$, $\theta_{cb} = $ 60$^\circ$ & 11.6 & 11.9 & 11.7 & 14.0 \\
\hline
8.$a_0 =$5, $y_0 = \lambda/4$, $\theta_{cb}= $ 80$^\circ$ & 9.3 & 9.3 & 9.8 & 16.6 \\
\hline
9.$a_0 =$5, $y_0 = \lambda/4$, $\theta_{cb}= $ 20$^\circ$ & 14.8 & 14.8 & 14.8 & 41.3 \\
\hline
10.$a_0 =$5, $y_0 = \lambda/4$, $\theta_{cb} = $ 10$^\circ$ & 48.7 & 48.3 & 49.0 & 62.5 \\
\hline
\end{tabular}
\end{center}
\caption{\label{table1}Summary of results for different cases.  Shown in the percentage of the total problem duration for which a given pusher is able to obtain convergence for $M \le$8.  The special cases of the Vay and Higuera-Cary pushers in the baseline case are noted by an asterisk.}
\end{table}

From Table \ref{table1} we find that the baseline case unfortunately represents a rather optimistic one from the point of view of numerically solving this problem.  In general we found that even the RK4 pusher was unable to produce converged solutions for more than 18\% of the problem duration.  Converged solutions over the full duration were only obtained by the RK4 solver in a couple of cases.  Also as the approach angle, $\theta_{cb}$, becomes very small, it is much easier to obtain convergence. 

All the leapfrog solvers perform less well than the RK4 pusher.  The differences between the three are usually rather small (again suggesting that the baseline case, happens to be a special case).  It therefore appears that, in general, the enhanced leapfrog solvers are not substantially better at the crossed beam problem than the Boris pusher.

\begin{figure}
\includegraphics[width=\columnwidth]{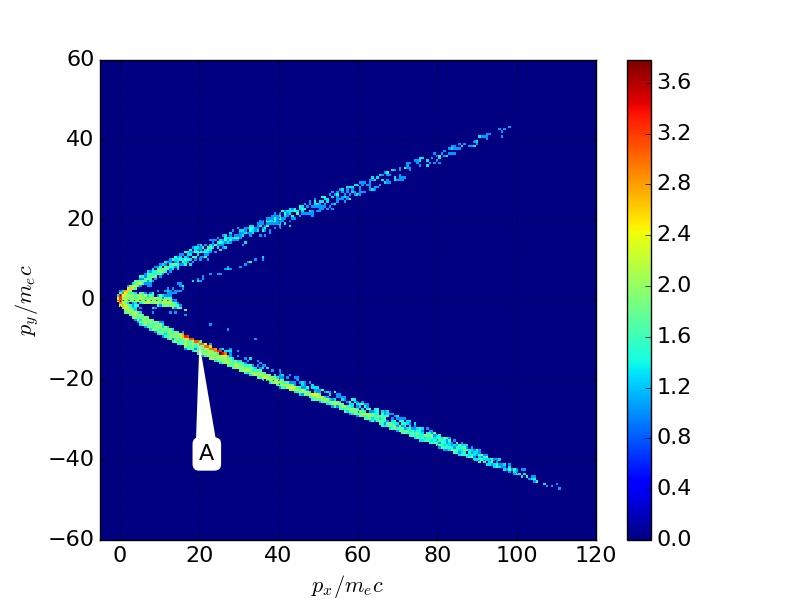}
\caption{\label{fig:figure9}Distribution at 450~fs of ensemble calculation (see text) for the case of the Boris solver. }
\end{figure}

\begin{figure}
\includegraphics[width=\columnwidth]{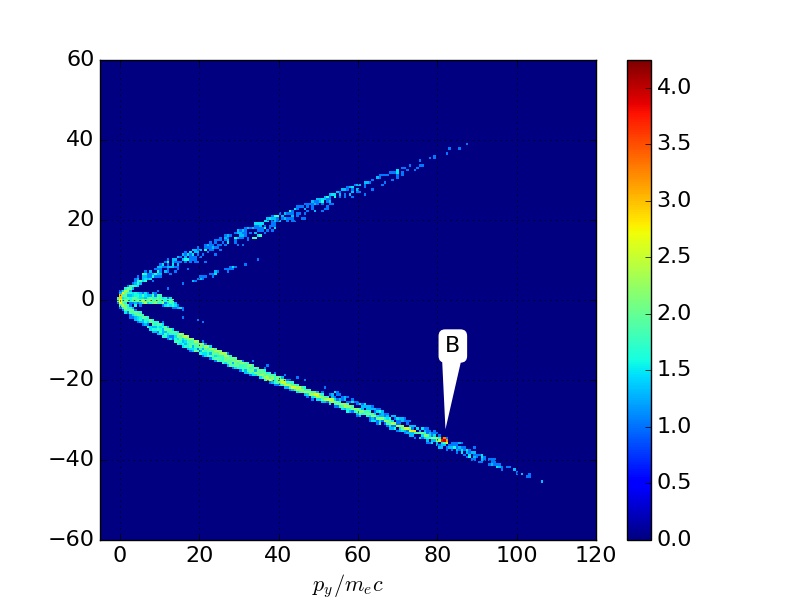}
\caption{\label{fig:figure10}Distribution at 450~fs of ensemble calculation (see text) for the case of the RK4 solver. }
\end{figure}

We have also examined the effect that the different solvers have on distributions arising from an ensemble of different initial conditions.  This was done for 10000 different particles initialized at rest with the initial $x$-position spanning -0.5 to +0.5$\lambda_L$ ($y_{init} = \lambda_L/4$).  The problem was run up to 450~fs with $M = $1.  Otherwise the problem corresponds to the baseline case.  We compared the distributions that arose from using the Boris and the RK4 solvers, which are shown in fig.s \ref{fig:figure9} and \ref{fig:figure10} respectively.


In the case of the RK4 solver we see that there is a very strong spike at high energy, denoted as 'B' in fig. \ref{fig:figure10}.  This feature is absent in fig. \ref{fig:figure9}, and instead we see a different feature denoated as 'A' in this figure.  Given that the strongest accumulations of particles are completely different for different solvers applied to the same ensemble/problem, we can conclude that the issues observed with single trajectories will lead to significant differences in particles distributions as well.

It therefore appears to be the case that the crossed beam problem presented here represents a far harder test than the single plane wave of single electron trajectory calculation.  To the best of the author's knowledge this is currently the hardest test case, at least specifically for laser-plasma studies, as the conventional particle pushers tested here are known to capable of producing fully converged solutions (for $M \le$ 8) over the full duration.  This is certainly the case for the single plane wave problem.  More importantly the results presented in Table \ref{table1} already indicate the most likely reason as to why this problem is so challenging : namely that the dynamics has become chaotic.  We see that, in the general case, a converged solution can only be obtained for a short period of time.  We also see that there are strange isolated cases where a full converged solution can be obtained.  The observation of these features motivated a more detailed study of the problem.

%###################################################
%##### Analysis Phase Two (Kavin)
%###################################################
\section{Parameter Scans with Advanced Algorithms}
\label{advanced}
In the second phase of this study, another class of solvers was used, namely the {\sc MATLAB} suite of ODE solvers.  In broad terms, applying these solvers to the problem lead to results similar to those presented in Sec.~\ref{standard}, with convergence only obtained over a limited period of time and for a small angle between the beams.  Out of the entire suite, ODE113 performed the best.  This solver is a variable-step, variable-order (VSVO) Adams-Bashforth-Moulton Predictor-Corrector solver of order 1 to 13.  It was found that convergence was reliably obtained when the angle between the beams was limited to no more than $\theta_{cb} =$ 30$^\circ$.  We have cross-checked the results obtained with ODE113 against the RK4 algorithm, and found good agreement between the two.  

In order to study the sensitivity to the initial conditions, parameters scans were then carried out by varying $\theta_{cb}$, $\phi_1$, and $\phi_2$.  For each set of initial conditions a calculation was run up to 200~fs.  Two outputs were recorded : (a) the ratio of the final displacement in $y$ to that in $x$ ($r_y/r_x$), and (b) the time at which the maximum $\gamma$ occurred ($\tau_{\gamma,{max}}$, normalized to the laser period).  Two types of scans were carried out, {\it coarse} and {\it fine}.  For the coarse scans, 100 points were used for each parameter over a large range : $\pm \pi$ for phases and 10--30$^\circ$ for $\theta_{cb}$.  For the fine scans, a fraction of each range was used and 200 points were then used for each parameter.  In all other respects, the calculations are the same as the baseline calculation described in Sec.~\ref{model}.  By moving from the analysis of Sec.~\ref{standard} where we looked at 10 cases to these parameter scans where we look at 10000-40000 cases per scan, we can obtain a much clearer idea of how sensitive this problem can be to the initial conditions.

\begin{figure}
\includegraphics[width=\columnwidth]{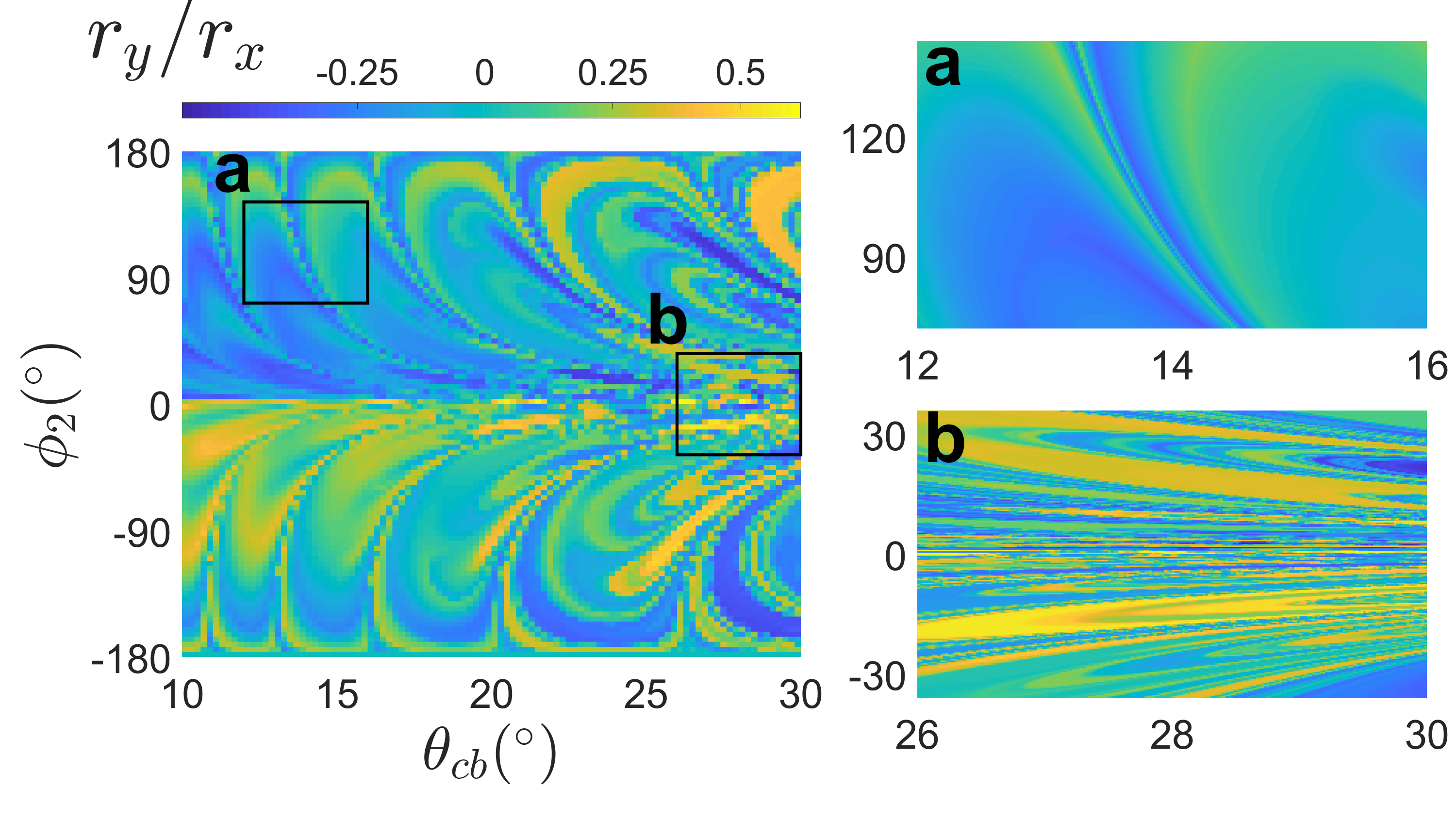}
\caption{\label{fig:figure6}Results from parameter scan over $\phi_2$ and $\theta_{cb}$ in terms of $r_y/r_x$.  Main plot is a coarse scan, with fine scans as sub-plots a and b. }
\end{figure}

\begin{figure}
\includegraphics[width=\columnwidth]{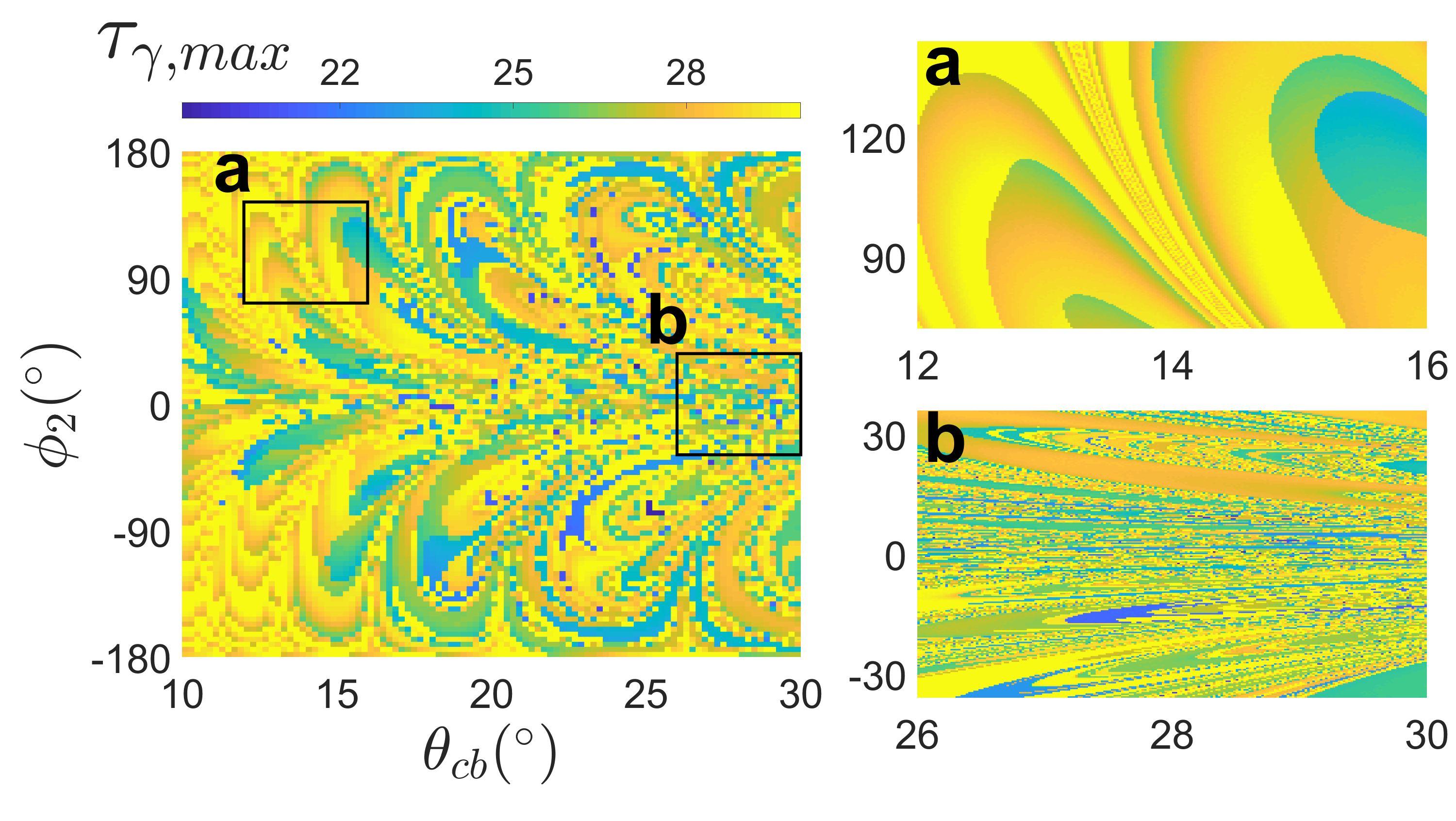}
\caption{\label{fig:figure7}Results from parameter scan over $\phi_2$ and $\theta_{cb}$ in terms of $\tau_{\gamma,{max}}$.  Main plot is a coarse scan, with fine scans as sub-plots a and b. }
\end{figure}

In Fig.s \ref{fig:figure6} and \ref{fig:figure7} we show the results from a coarse parameter scan of $\theta_{cb}$ and $\phi_2$ with $\phi_1$ held fixed at 0$^{\circ}$.  The sub-figures show plots of fine parameter scans in the regions indicated.

What we observe from these extensive parameters scans is that the parameter space appears to consist of two types of regions.  There are regions where the results of the calculations vary (relatively) slowly and smoothly as the initial conditions are changed.  Examples of these are shown in the `a' sub-figures in both fig.~\ref{fig:figure6}  and fig.~\ref{fig:figure7}.  There are also regions where small variations of the initial conditions leads to gross changes in the results including rapid changes in sign.  Examples of these sub-regions are shown in the `b' sub-figures in both fig.~\ref{fig:figure6}  and fig.~\ref{fig:figure7}.  

\begin{figure}
\includegraphics[width=\columnwidth]{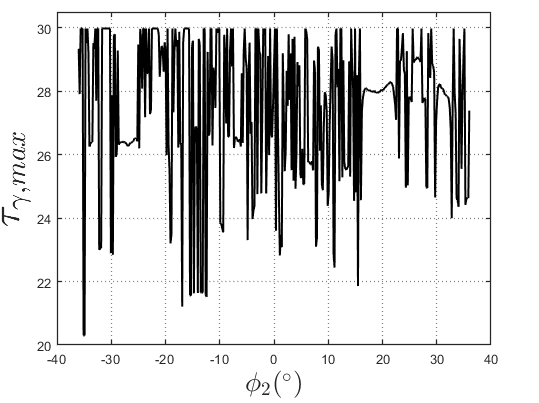}
\caption{\label{fig:figure8}Line-out of Fig.\ref{fig:figure7}(b) along $\theta_{cb} =$ 28$^\circ$}
\end{figure}

As we were observing strong point-to-point changes in fig.~\ref{fig:figure7}(b) along $\theta_{cb} = $ 28$^\circ$ we repeated this set of calculations at twice the resolution in $\phi_2$ (i.e. now with 400 points in $\phi_2$ across the `fine' range).  The results are shown in fig.~\ref{fig:figure8}.  It can be seen that there is no improvement in terms of being able to `resolve' the detail in this region.

We can summarize the results from this second phase of the study as follows : (i) we have done an extensive parameter scan of the initial conditions / problem parameters using an advanced ODE solver, (ii) this reveals regions in parameter space that are very sensitive to the initial conditions / problem parameters, (iii) we are not able to `resolve' this sensitivity by successively refining the set of points over which we scan.  These observations suggest that we are  actually looking at a system that exhibits chaotic dynamics, as we expected from earlier studies.

%###################################################
%##### Conclusions
%###################################################
\section{Conclusions}
In this paper we have examined an apparently simple model problem in relativistic single electron motion relevant to ultra-intense laser-plasma interactions, involving two obliquely intersecting plane wave-packets.  The findings for this model problem, which are presented herein can be summarized as follows:

\begin{enumerate}
\item{Under a wide range of conditions converged solutions cannot be obtained for a 1~ps period using a wide range of different solvers including the Boris method, 4th order Runge-Kutta, and the {\sc MATLAB} suite of ODE solvers.}
\item{Converged solutions appear to occur in isolated ranges of problem parameters.}
\item{Converged solutions can, in general, only be obtained over quite short durations, especially compared to benchmarks such as the single plane-wave problem where this is not an issue.}
\item{When extensive parameter scans are carried out across initial conditions / problem parameters, it is found that regions in parameter space exist where there is a very high degree of sensitivity to these initial conditions (or problem parameters).}
\item{Progressively increasing the resolution of these sensitive regions does not lead to any improved resolution of the highly sensitive region.}
\end{enumerate} 

Our findings have, in the authors' view, two main consequences.  Firstly, great care needs to be taken when using PIC codes to study laser-plasma interactions.  Prior to this study it was generally assumed that algorithms such as the Boris pusher would produce reasonably accurate results irrespective of the field configuration under consideration.  In light of this study, we no longer think this can be assumed.  We suggest that PIC simulations are accompanied by complementary studies of the single particle motion to ensure that converged orbits can be obtained.

Secondly, these findings suggest that the root cause of both the issues of convergence and the sensitivity to initial conditions is at the very least indicative of extreme nonlinearity, but it quite strongly suggests that the dynamics of this problem are {\it chaotic}.  This is entirely consistent with earlier studies \cite{z.-m.sheng_stochastic_2002,sheng_efficient_2004,meyer-ter-vehn_electron_1999}, however these results now indicate that it is quite easy for the Lyapunov time to become sufficiently short that numerical integration is inhibited.  This would explain the very limited ability of nearly all methods to obtain converged solutions, and it also explains the very high sensitivity to initial conditions.  We do not claim to provide any rigorous proof that the dynamics of this system are chaotic, only to submit the results of numerical calculations that show that this might be the case, and that further investigation should be carried out.  We do however draw the attention of the reader to earlier studies where such detailed analysis was carried out \cite{mendonca_1983}.  If this simple model problem is indeed shown to have chaotic dynamics then this could have quite profound implications for the field of ultra-intense laser-plasma interactions, as it would then imply that a number of laser-target configurations where there are interfering laser fields would have the potential for chaotic dynamics.

\section*{Acknowledgements}
The work of K.T. and A.V.A  was supported by the National Science Foundation (PHY 1632777). K.W. was supported in part by the DOE Office of Science under Grant No. DE-SC0018312 and in part by the DOE Computational Science Graduate Fellowship under Grant No. DE-FG02-97ER25308. 

% \bibliography{zot_bib}
%merlin.mbs aipnum4-1.bst 2010-07-25 4.21a (PWD, AO, DPC) hacked
%Control: key (0)
%Control: author (8) initials jnrlst
%Control: editor formatted (1) identically to author
%Control: production of article title (-1) disabled
%Control: page (0) single
%Control: year (1) truncated
%Control: production of eprint (0) enabled
%

\end{document}